\documentclass[journal]{IEEEtran}
\usepackage[keeplastbox]{flushend}

\usepackage[noadjust]{cite}

\ifCLASSINFOpdf
\usepackage[pdftex]{graphicx}
\else
\usepackage[dvips]{graphicx}
\fi
\usepackage{amsmath}
\usepackage{amssymb}
\usepackage{algorithm}
\usepackage{algorithmic}

\usepackage{array}

\ifCLASSOPTIONcompsoc
  \usepackage[caption=false,font=normalsize,labelfont=sf,textfont=sf]{subfig}
\else
  \usepackage[caption=false,font=footnotesize]{subfig}
\fi
\begin{document}
%
\title{Sphere Bounding Scheme for Probabilistic\\Robust Constructive Interference Precoding\\in MISO Downlink Transmission}
%
%
%

\author{Yuning~You
        and
        Gangming~Lv,~\IEEEmembership{Member,~IEEE}
\thanks{Y. You and G. Lv are with the Department of Information and Communications Engineering, Xi'an Jiaotong University, Xi'an, Shaanxi, 710049, P.R. China (e-mail: yuning.you@foxmail.com; gmlv@mail.xjtu.edu.cn).}}
\maketitle

\begin{abstract}
In this letter, we propose a sphere bounding scheme for probabilistic robust constructive interference (CI) power minimizing precoding, to address the imperfect channel state information (CSI) caused by the channel error (CE), which satisfies the known distribution in single-cell multiuser multiple-input single-output (MISO) downlink transmission. In the proposed scheme, we transform the probabilistic quality of service (QoS) constraints into tractable sphere bounding second-order cone (SOC) constraints through taking two-step tightening, and then we model tightened CI max-min signal-to-noise ratio (SNR) precoding, proving that its lower bound can be solved through tightened CI power minimizing precoding. Besides, in tightened CI power minimizing precoding, we propose the relaxation iteration to relax the connect probability requirement. Finally, we analyze the complexity of our proposed scheme. Numerical results show that our proposed schemes perform well in the satisfaction of the connect probability requirement, resulting in lower symbol error rate (SER) and higher transmit power.
\end{abstract}


%
\IEEEpeerreviewmaketitle

\section{Introduction}
%
%
%
%
\IEEEPARstart {I}{n} recent years, precoding has been done a prolific research on for its guarantee of breaking the throughput gridlock of many communication transmissions \cite{malodeh0318}, e.g. MISO downlink transmission. In general, it assumes that perfect CSI is estimated for precoding, which cannot be true in all practical scenarios. Therefore the robust design tackling imperfect CSI is required. For traditional precoding which utilizes CSI without data information (DI), the robust design is well studied in literatures, e.g. the worse-case design \cite{jwang0809} assuming CE lies in a bounded set and ensuring the QoS constraints satisfied for bounded CE, and the probabilistic design \cite{kwang1114} assuming the distribution CE satisfied, and promising the QoS constraints satisfied for a certain probability.

With the development of CI precoding \cite{cmasouros0715} utilizing both CSI and DI to exploit the interference instead of eliminate it as traditional precoding, CI precoding shows great advantage to traditional precoding to guarantee interference-free communication in spite of its higher switching rate. Nevertheless, there is room to develop the robust design for CI precoding. In \cite{cmasouros0715}, the worse-case design is proposed, bounding CE and ensuring the QoS constraints satisfied for bounded CE. In \cite{dkwon0416}, unpredictable CE caused by quantization is studied and also modeled as bounded. In \cite{ahaqiqatnejad0018}, the probabilistic design is proposed for known distribution CE, promising the QoS constraints satisfied for a certain probability, but its derivation is incorrect, resulting in the unstable performance. To the best of our knowledge, there is no other probabilistic design for CI precoding, which is the motivation of this letter. The main challenge of the probabilistic design lies in its computational intractability resulting from the probabilistic QoS constraints, in contrast to the efficient solvability of the non-robust design.

In this letter, we propose a sphere bounding scheme for probabilistic robust CI power minimizing precoding with known distribution of CE, through tightening the probabilistic QoS constraints into tractable sphere bounding SOC constraints, and then we prove the lower bound of the tightened CI max-min SNR precoding can be solved through tightened CI power minimizing precoding. In addition, in tightened CI power minimizing precoding, we propose the relaxation iteration for the connect probability requirement. Finally, we analyze the complexity of our proposed schemes. Numerical results are provided, demonstrating connect probability, SER and transmit power against SNR target.

\emph{Notation:} $\boldsymbol{0}$ is the vector or matrix of all zeros with the size determined by context, $\boldsymbol{I}_K\in \boldsymbol{R}^{K\times K}$ is the identity matrix, $\boldsymbol{\zeta}_K\sim \mathcal{CN}(\boldsymbol{0}, \boldsymbol{I}_K)$, $\boldsymbol{\varepsilon}_K\sim \mathcal{N}(\boldsymbol{0}, \boldsymbol{I}_K)$, and constr. is the abbreviation of constraint.

\section{System Model}

We consider the single-cell multiuser MISO downlink transmission, where there is a base station (BS) equipped with $M$ transmit antennas, sending data to $N$ single-received-antenna users. We assume the adopted modulation method is MPSK, $d_i\in \{{\rm exp}({\rm j}2\theta \times 0),\dots ,{\rm exp}({\rm j}2\theta \times (\mathcal{M} -1))\}$, $i\in N$, is the sent data for the $i$th user, where $\theta =\pi /\mathcal{M}$ and $\mathcal{M}$ is the modulation order, and the channel between transmitter and receiver is block quasi-static and flat fading. Thus the received signal at the $i$th user can be written as
\begin{equation}
y_i = \boldsymbol{h}_i^{\rm T}\boldsymbol{x} + z_i, \label{e1}
\end{equation}

\noindent where $\boldsymbol{x}\in \mathcal{C}^{M\times 1}$ is the transmitted signal vector from BS, $\boldsymbol{h}_i\in \mathcal{C}^{M\times 1}$ is the independent CSI vector between BS and the $i$th user, and $z_i\sim \mathcal{CN}(0, \sigma_{z_i}^2 )$ is the independent received noise. Transmit power at BS is defined as $P\triangleq \| \boldsymbol{x}\|^2$, and SNR at the $i$th user is defined as $\gamma_i\triangleq \| \boldsymbol{h}_i^{\rm T}\boldsymbol{x}\|^2/{\sigma_{z_i}^2}$. When considering imperfect CSI, $\boldsymbol{h}_i$ can be modeled as
\begin{equation}
\boldsymbol{h}_i = \boldsymbol{h}_{{\rm est} ,i} + \boldsymbol{e}_i, \label{e2}
\end{equation}

\noindent where $\boldsymbol{h}_{{\rm est} ,i}$ is the estimated imperfect CSI vector, $\boldsymbol{e}_i\sim \mathcal{CN} (\boldsymbol{0}, \boldsymbol{\Sigma}_{\boldsymbol{e}_i})$ is the independent CE vector, and $\boldsymbol{\Sigma}_{\boldsymbol{e}_i}={\rm diag}(\sigma_{\boldsymbol{e}_i, 1}^2, \dots , \sigma_{\boldsymbol{e}_i, M}^2)$.

\subsection{CI Definition}

Referring to \cite{cmasouros0715, ahaqiqatnejad0418}, the definition of distance preserve CI for MPSK can be stated as followed.

\newtheorem{definition}{Definition}
\begin{definition}
The $i$th user, associated with $\boldsymbol{h}_i$ and $d_i$, is said to receive CI from $\boldsymbol{x}$ under the SNR requirement $\gamma_i \ge \hat{\gamma}_i$ where $\hat{\gamma}_i \ge 0$, if and only if the following inequality holds
\begin{equation}
| {\rm Im}(d_i^*\boldsymbol{h}_i^{\rm T}\boldsymbol{x}) |/{{\rm tan}\theta} \le {\rm Re}(d_i^*\boldsymbol{h}_i^{\rm T}\boldsymbol{x}) - \sqrt{\hat{\gamma}_i}\sigma_{z_i}. \label{e3}
\end{equation}
\end{definition}

\subsection{CI Power Minimizing Precoding with Imperfect CSI and its Real-Valued Representation}
CI power minimizing precoding \cite{cmasouros0715} is a conventional CI precoding, aimed at minimizing transmit power with \eqref{e3} satisfied at all users. When considering imperfect CSI in precoding, utilizing the fact that $d_i^*$ is specific, we have $d_i^*\boldsymbol{e}_i = \boldsymbol{\Sigma}_{\boldsymbol{e}_i}^{1/2}\boldsymbol{\zeta}_M$. Thus, with \eqref{e2}, constr. \eqref{e3} in CI power minimizing precoding can be rewritten as
\begin{align}
& | {\rm Im}((d_i^*\boldsymbol{h}_{{\rm est},i} + \boldsymbol{\Sigma}_{\boldsymbol{e}_i}^{1/2}\boldsymbol{\zeta}_M)^{\rm T}\boldsymbol{x}) |/{{\rm tan}\theta} \notag
\\ \le {} & {\rm Re}((d_i^*\boldsymbol{h}_{{\rm est},i} + \boldsymbol{\Sigma}_{\boldsymbol{e}_i}^{1/2}\boldsymbol{\zeta}_M)^{\rm T}\boldsymbol{x}) - \sqrt{\hat{\gamma}_i}\sigma_{z_i}. \label{e4}
\end{align}

\noindent For convenience, we will transform the complex representation of constr. \eqref{e4} into a real-valued representation. Let $\tilde{\boldsymbol{h}}_i = [{\rm Re}(d_i^*\boldsymbol{h}_{{\rm est},i}); {\rm Im}(d_i^*\boldsymbol{h}_{{\rm est},i})]$, $\tilde{\boldsymbol{\Sigma}}_{\boldsymbol{e}_i} = [\boldsymbol{\Sigma}_{\boldsymbol{e}_i}, \boldsymbol{0}; \boldsymbol{0}, \boldsymbol{\Sigma}_{\boldsymbol{e}_i}]/2$, $\tilde{\boldsymbol{x}} = [{\rm Re}(\boldsymbol{x}); {\rm Im}(\boldsymbol{x})]$, $\boldsymbol{A} = [\boldsymbol{I}_M, \boldsymbol{0}; \boldsymbol{0}, -\boldsymbol{I}_M]$, and $\boldsymbol{B} = [\boldsymbol{0}, \boldsymbol{I}_M; \boldsymbol{I}_M, \boldsymbol{0}]$, due to $[{\rm Re}(\boldsymbol{\Sigma}_{\boldsymbol{e}_i}^{1/2}\boldsymbol{\zeta}_{M}); {\rm Im}(\boldsymbol{\Sigma}_{\boldsymbol{e}_i}^{1/2}\boldsymbol{\zeta}_{M})] = \tilde{\boldsymbol{\Sigma}}_{\boldsymbol{e}_i}^{1/2}\boldsymbol{\varepsilon}_{2M}$, we can equivalently transform constr. \eqref{e4} into
\begin{align}
&|(\tilde{\boldsymbol{h}}_i + \tilde{\boldsymbol{\Sigma}}_{\boldsymbol{e}_i}^{1/2}\boldsymbol{\varepsilon}_{2M})^{\rm T}\boldsymbol{B}\tilde{\boldsymbol{x}}|/{{\rm tan}\theta} \notag
\\ \le {} & (\tilde{\boldsymbol{h}}_i + \tilde{\boldsymbol{\Sigma}}_{\boldsymbol{e}_i}^{1/2}\boldsymbol{\varepsilon}_{2M})^{\rm T}\boldsymbol{A}\tilde{\boldsymbol{x}} - \sqrt{\hat{\gamma}_i}\sigma_{z_i}. \label{e5}
\end{align}

\noindent The real-valued representation of CI power minimizing precoding with imperfect CSI can be presented as
\begin{subequations}\label{e6}
\begin{gather}{\rm min}_{\tilde{\boldsymbol{x}}}\, \|\tilde{\boldsymbol{x}}\|^2 \label{e6_1}\\
\text{s.t.}\quad \text{constr. } \eqref{e5}, i\in N. \label{e6_2} \end{gather}
\end{subequations}

\section{Probabilistic Design\\with Sphere Bounding Scheme}

For the existence $\boldsymbol{\varepsilon}_{2M}$ in constr. \eqref{e6_2}, optimization problem \eqref{e6} is intractable. Rather than addressing constr. \eqref{e6_2} directly, we address connect probability, which at the $i$th user is defined as $p_i \triangleq {\rm Prob}\{\text{constr. } \eqref{e5} \text{ holds}\}$, formulating the probabilistic design as
\begin{subequations}\label{e7}
\begin{gather}{\rm min}_{\tilde{\boldsymbol{x}}}\, \|\tilde{\boldsymbol{x}}\|^2 \label{e7_1} \\
\text{s.t.}\quad p_i \ge \hat{p}_i, i\in N, \label{e7_2} \end{gather}
\end{subequations}

\noindent where $0 \le \hat{p}_i \le 1$ is the connect probability requirement at the $i$th user. Aimed at tackling optimization problem \eqref{e7}, in this section, we propose a sphere bounding scheme through taking two-step tightening. For ease of composition, we omit the user index in the following derivation.

\subsection{Sphere Bounding Scheme}

In optimization problem \eqref{e7}, it is still difficult to tackle constr. \eqref{e7_2}, and therefore, it is necessary to tighten it properly. We will analyze constr. \eqref{e7_2} directly, and then take a two-step tightening for it.

In constr. \eqref{e7_2}, noticing that there is absolute value in connect probability, we can rewrite connect probability as
\begin{equation}
p = {\rm Prob}\{v_1 \ge w_1, v_2 \ge w_2\},
 \label{e8}
\end{equation}

\noindent where
$v_1 = \boldsymbol{\varepsilon}_{2M}^{\rm T}\tilde{\boldsymbol{\Sigma}}_{\boldsymbol{e}}^{1/2}(\boldsymbol{A}-\boldsymbol{B}/{{\rm tan}\theta})\tilde{\boldsymbol{x}}$,
$w_1 = \tilde{\boldsymbol{h}}^{\rm T}(-\boldsymbol{A}+\boldsymbol{B}/{{\rm tan}\theta})\tilde{\boldsymbol{x}} +  \sqrt{\hat{\gamma}}\sigma_{z}$,
$v_2 = \boldsymbol{\varepsilon}_{2M}^{\rm T}\tilde{\boldsymbol{\Sigma}}_{\boldsymbol{e}}^{1/2}(\boldsymbol{A}+\boldsymbol{B}/{{\rm tan}\theta})\tilde{\boldsymbol{x}}$,
and
$w_2 = \tilde{\boldsymbol{h}}^{\rm T}(-\boldsymbol{A}-\boldsymbol{B}/{{\rm tan}\theta})\tilde{\boldsymbol{x}} +  \sqrt{\hat{\gamma}}\sigma_{z}$.
For any given $\tilde{\boldsymbol{h}}$ and $\tilde{\boldsymbol{x}}$, $w_1$ and $w_2$ can be taken as two constants, while it is intuitive to verify that the random variable $(v_1, v_2)^{\rm T}$ satisfies the bivariate normal distribution as
\begin{equation}
(v_1, v_2)^{\rm T}\sim \mathcal{N}(0, 0, \sigma_{v_1}^2, \sigma_{v_2}^2, \rho),
 \label{e9}
\end{equation}

\noindent where
$\sigma_{v_1}^2 = \|\tilde{\boldsymbol{\Sigma}}_{\boldsymbol{e}}^{1/2}(\boldsymbol{A}-\boldsymbol{B}/{{\rm tan}\theta})\tilde{\boldsymbol{x}}\|^2$,
$\sigma_{v_2}^2 = \|\tilde{\boldsymbol{\Sigma}}_{\boldsymbol{e}}^{1/2}(\boldsymbol{A}+\boldsymbol{B}/{{\rm tan}\theta})\tilde{\boldsymbol{x}}\|^2$,
and
$\rho = \tilde{\boldsymbol{x}}^{\rm T}(\boldsymbol{A}-\boldsymbol{B}/{{\rm tan}\theta})^{\rm T}\tilde{\boldsymbol{\Sigma}}_{\boldsymbol{e}}(\boldsymbol{A}+\boldsymbol{B}/{{\rm tan}\theta})\tilde{\boldsymbol{x}}/{(\sigma_{v_1}\sigma_{v_2})}$.
Therefore, we can calculate connect probability in \eqref{e8} with the bivariate normal integral, however which can only be presented as the infinite series consisting of the special function \cite{hafayed0114}. In \cite{ahaqiqatnejad0018}, $(v_1, v_2)^{\rm T}$ is decorrelated, incorrectly transforming the bivariate normal integral into an equivalent simplified representation, resulting in the unstable performance. Here we will take another route to address connect probability in \eqref{e8}, transform it into
\begin{align}
p ={} & 1 - {\rm Prob}\{v_1 \le w_1\} - {\rm Prob}\{v_2 \le w_2\} \notag
\\ & + {\rm Prob}\{v_1 \le w_1, v_2 \le w_2\} \notag
\\={} & -({\rm erf}(w_1/(\sqrt{2}\sigma_{v_1})) + {\rm erf}(w_2/(\sqrt{2}\sigma_{v_2})))/2 \notag
\\ & + {\rm Prob}\{v_1 \le w_1, v_2 \le w_2\}
 \label{e10}
\end{align}

\noindent Since ${\rm Prob}\{v_1 \le w_1, v_2 \le w_2\} \ge 0$, we take the first-step tightening for constr. \eqref{e7_2} as
\begin{equation}
-({\rm erf}(w_1/(\sqrt{2}\sigma_{v_1})) + {\rm erf}(w_2/(\sqrt{2}\sigma_{v_2})))/2 \ge \hat{p},
 \label{e11}
\end{equation}
\noindent which is the necessary condition for constr. \eqref{e7_2}. In the first-step tightening, we tighten the value of ${\rm Prob}\{v_1 \le w_1, v_2 \le w_2\}$, simplifying the bivariate normal integral into the sum of two error functions. Thus, the tightness performance is dependent on the value of ${\rm Prob}\{v_1 \le w_1, v_2 \le w_2\}$. Simulation shows that, with the higher connect probability requirement and modulation order, we will have better tightness performance in the first-step tightening.

Following constr. \eqref{e11}, we can see that it is still not a tractable constr., and therefore we need to take a second-step tightening to transform it into a tractable one. Let $\mu_1 = w_1/(\sqrt{2}\sigma_{v_1})$, and $\mu_2 = w_2/(\sqrt{2}\sigma_{v_2})$, we define the function $g(\mu_1, \mu_2) \triangleq -({\rm erf}(\mu_1) + {\rm erf}(\mu_2))/2$. Constr. \eqref{e11} can be rewritten as
\begin{equation}
g(\mu_1, \mu_2) \ge \hat{p}.
 \label{e12}
\end{equation}

\noindent The monotonicity of $g(\mu_1, \mu_2)$ can be easily derived as
\begin{gather}
\frac{\partial g(\mu_1, \mu_2)}{\partial \mu_1} = -\dfrac{1}{2} \times \dfrac{d {\rm erf}(\mu_1)}{d \mu_1} \le 0, \notag\\
\frac{\partial g(\mu_1, \mu_2)}{\partial \mu_2} = -\dfrac{1}{2} \times \dfrac{d {\rm erf}(\mu_2)}{d \mu_2} \le 0.
\label{e13}
\end{gather}

\noindent With the monotonicity, we can take the second-step tightening, giving the necessary condition for constr. \eqref{e12} as
\begin{equation}
\mu_1 \le \lambda_1, \mu_2 \le \lambda_2, g(\lambda_1, \lambda_2) = \hat{p}.
\label{e14}
\end{equation}

\begin{figure}[!t]
\centering
\includegraphics[width=2.3in]{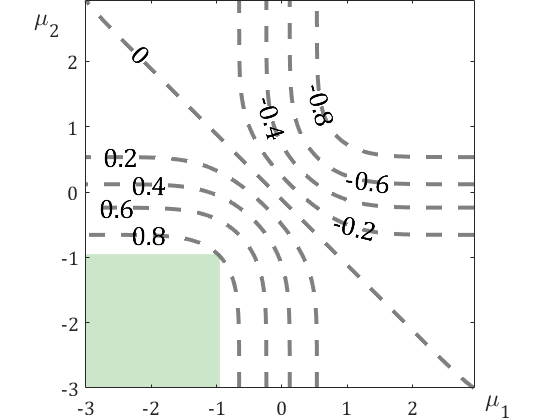}
\caption{The contour plot of $g(\mu_1, \mu_2)$. The green region is the tightened region for $\{(\mu_1, \mu_2)|\,g(\mu_1, \mu_2) \ge 0.8\}$.}
\label{f1}
\end{figure}

To have a straight insight of the second-step tightening, we give the contour plot of $g(\mu_1, \mu_2)$, illustrating the original region and the tightened region in Fig. \ref{f1}. The original region as constr. \eqref{e12} is shaped as a rounded rectangle region wrapped by the contour line, and the tightened region as constr. \eqref{e14} is a rectangle region used to approximating the rounded rectangle region (e.g. for the original region $\{(\mu_1, \mu_2)|\, g(\mu_1, \mu_2) \ge 0.8\}$, its tightened region can be presented as the green region in Fig. \ref{f1}). Simulation shows that, with the higher connect probability requirement, we will have better tightness performance in the second-step tightening.

Without other consideration, we require $\lambda_1 = \lambda_2$ in constr. \eqref{e14}. With the definition of $w_1$, $w_2$, $\sigma_{v_1}$ and $\sigma_{v_2}$, we can rewrite constr. \eqref{e14} as
\begin{align}
&r\|\tilde{\boldsymbol{\Sigma}}_{\boldsymbol{e}}^{1/2}(\boldsymbol{A}-\boldsymbol{B}/{{\rm tan}\theta})\tilde{\boldsymbol{x}}\| \le \tilde{\boldsymbol{h}}^{\rm T}(\boldsymbol{A}-\boldsymbol{B}/{{\rm tan}\theta})\tilde{\boldsymbol{x}} - \sqrt{\hat{\gamma}}\sigma_{z}, \notag \\
&r\|\tilde{\boldsymbol{\Sigma}}_{\boldsymbol{e}}^{1/2}(\boldsymbol{A}+\boldsymbol{B}/{{\rm tan}\theta})\tilde{\boldsymbol{x}}\| \le \tilde{\boldsymbol{h}}^{\rm T}(\boldsymbol{A}+\boldsymbol{B}/{{\rm tan}\theta})\tilde{\boldsymbol{x}} - \sqrt{\hat{\gamma}}\sigma_{z}, \notag \\
&r = \sqrt{2}{\rm erf}^{-1}(\hat{p}).
\label{e15}
\end{align}

\noindent This is the final tightened constr. for constr. \eqref{e7_2}. Noticing that constr. \eqref{e15} takes the same form as the constr. in the worse-case design \cite{cmasouros0715}, bounding CE in a sphere region with the pre-specified radius $r$, in contrast to constr. \eqref{e15}, controlling the radius $r$ with the connect probability requirement. The sphere bounding scheme for probabilistic robust CI power minimizing precoding can be finally expressed as
\begin{subequations}\label{e16}
\begin{gather}{\rm min}_{\tilde{\boldsymbol{x}}}\, \|\tilde{\boldsymbol{x}}\|^2 \label{e16_1} \\
\text{s.t.}\quad \text{constr. } \eqref{e15}, i\in N, \label{e16_2} \end{gather}
\end{subequations}

\noindent which is a standard SOC program and therefore can be efficiently solved. Since tightened constr. \eqref{e16_2} is stronger than constr. \eqref{e7_2}, the solution of optimization problem \eqref{e16} is the upper bound of the solution of optimization problem \eqref{e7}.

\subsection{Sphere Bounding Scheme for CI Max-Min SNR Precoding}

Following the modeling and tightening of probabilistic robust CI power minimizing precoding as optimization problem \eqref{e16}, we can model the sphere bounding scheme for probabilistic robust CI max-min SNR precoding as
\begin{subequations}\label{e17}
\begin{gather}  {\rm max}_{\tilde{\boldsymbol{x}}}\, {\rm min}\{\gamma,i\in N\} \label{e17_1} \\
\text{s.t.}\quad  \|\tilde{\boldsymbol{x}}\|^2 \le \hat{P}, \label{e17_2} \\
 \hat{\gamma} = \gamma, \text{constr. } \eqref{e16_2}, i\in N, \label{e17_3} \end{gather}
\end{subequations}

\noindent where $\hat{P}$ is the transmit power requirement.

\newtheorem{lemma}{Lemma}
\begin{lemma}\label{l1}
The lower bound of the solution of optimization problem \eqref{e17} can be solved through optimization problem \eqref{e16}.
\end{lemma}

\begin{IEEEproof}
See Appendix A for the proof.
\end{IEEEproof}

\subsection{Relaxation Iteration}

Recalling the definition of connect probability, in optimization problem \eqref{e16} we define actual connect probability $p_{act} \triangleq {\rm Prob}\{\text{constr. } \eqref{e5} \text{ holds} |\, \text{constr. } \eqref{e16_2} \text{ is satisfied}\}$. Due to the tightened constr. in optimization problem \eqref{e16}, we have
\begin{equation}
p_{act} \ge \hat{p}.
\label{e18}
\end{equation}

\noindent With the connect probability requirement $\hat{p}$, in practice we can relax constr. \eqref{e16_2} by setting a lower connect probability requirement in optimization problem \eqref{e16}. Since after the solution of optimization problem \eqref{e16}, we always can calculate actual connect probability with numerical integration method or Monte Carlo method, we consider to take a relaxation iteration to adjust the connect probability requirement. Let the probability difference $\Delta p = p_{act} - \hat{p}$, and $\hat{p}'$ be the adjusted connect probability requirement, for the $l$th iteration we calculate $\Delta p^{(l)}$ and $\hat{p}'^{(l)}$ as
\begin{equation}
\Delta p^{(l)} = p_{act}^{(l)} - \hat{p},
\hat{p}'^{(l)} = \hat{p}'^{(l-1)} + \eta \Delta p^{(l)},
\label{e19}
\end{equation}

\noindent where the step length $\eta > 0$, and we set $\hat{p}'^{(0)} = \hat{p}$. In the relaxation iteration, we update the adjusted connect probability requirement, and then solve optimization problem \eqref{e16} until the termination condition is satisfied. The detailed process of the iterative sphere bounding scheme is given following.
\begin{algorithm}[!h]
\caption{Iterative Sphere Bounding Scheme}
\label{a1}
\begin{algorithmic}
\STATE Set $\eta > 0$, $p_{act}^{(0)} \leftarrow 0$, $\Delta p^{(0)} \leftarrow 0$, $\hat{p}'^{(0)} \leftarrow \hat{p}$, $l \leftarrow 0$, $\delta > 0$
\REPEAT
\STATE $l \leftarrow l+1$
\STATE Solve optimization problem \eqref{e16} with $\hat{p}'^{(l-1)}$
\STATE Calculate $p_{act}^{(l)}$, $\Delta p^{(l)} \leftarrow p_{act}^{(l)} - \hat{p}$, $\hat{p}'^{(l)} \leftarrow \hat{p}'^{(l-1)} + \eta \Delta p^{(l)}$
\UNTIL{$|\Delta p^{(l)}| \le \delta$}
\end{algorithmic}
\end{algorithm}
\begin{figure*}[!t]
\centering
\subfloat[Connect probability vs SNR requirement (dB).]{\includegraphics[width=2.3in]{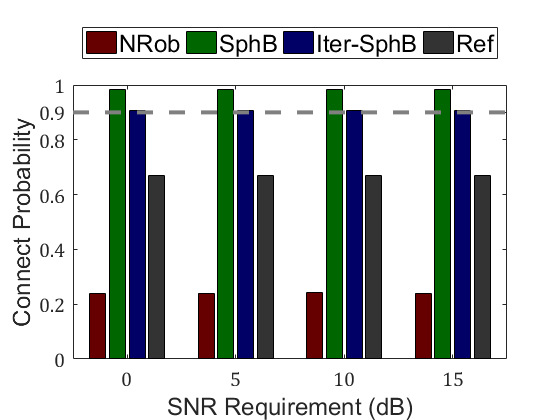}
\label{f2a}}
\hfil
\subfloat[SER vs SNR requirement (dB).]{\includegraphics[width=2.3in]{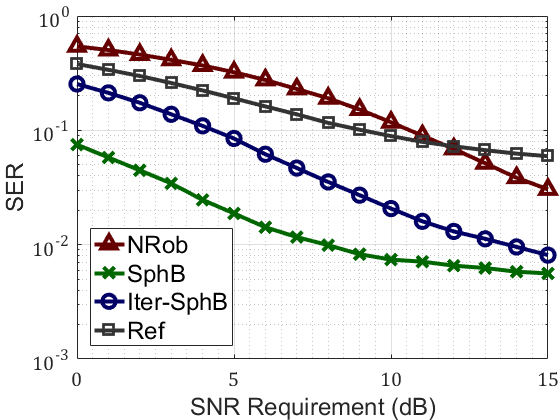}
\label{f2b}}
\hfil
\subfloat[Transmit power vs SNR requirement (dB).]{\includegraphics[width=2.3in]{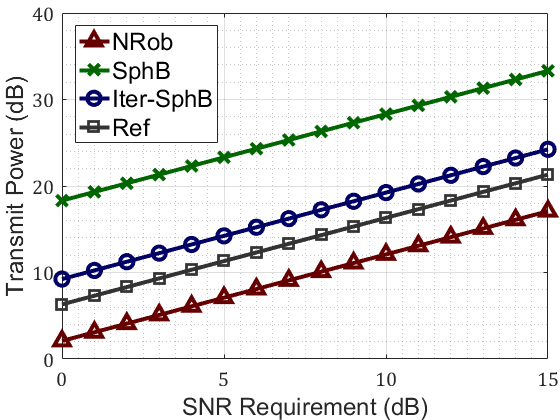}
\label{f2c}}
\caption{Numerical results.}
\label{f2}
\end{figure*}

\subsection{Complexity Analyses}

In the sphere bounding scheme, optimization problem \eqref{e16} has $2N$ SOC constraints of dimension $M+1$. According to \cite{kwang1114, abental0001}, given an $\epsilon > 0$, when it reaches an $\epsilon$-optimal solution, its iteration complexity order can be calculated as $2\sqrt{N}\times {\rm ln}(1/\epsilon)$, and its per-iteration computation cost order can be calculated as $\mathcal{O}(NM^2)\times 2N(M+1)^2+\mathcal{O}^3(NM^2)$, and therefore its complexity order is $2\sqrt{N}\times {\rm ln}(1/\epsilon)\times \mathcal{O}(NM^2)\times (2N(M+1)^2 + \mathcal{O}^2(NM^2))$. In the iterative sphere bounding scheme, assuming the integral complexity order and the iteration complexity order is $o_1$ and $o_2$, its complexity order can be calculated as $(2\sqrt{N}\times {\rm ln}(1/\epsilon)\times \mathcal{O}(NM^2)\times (2N(M+1)^2 + \mathcal{O}^2(NM^2)) + o_1)\times o_2$.

\section{Numerical Results}

To the best of our knowledge, there is no other probabilistic design for CI precoding, and therefore in simulation we will compare the sphere bounding scheme (SphB) and the iterative sphere bounding scheme (Iter-SphB) with the conventional non-robust precoding (NRob) \cite{cmasouros0715} and the reference probabilistic design (Ref) \cite{ahaqiqatnejad0418}. Under 8PSK modulation, we set $M=N=4$, $\sigma_{z_1}^2=\dots=\sigma_{z_N}^2=1$, $\boldsymbol{h}_{{\rm est},1},\dots,\boldsymbol{h}_{{\rm est},N}\sim \mathcal{CN}(\boldsymbol{0},\boldsymbol{I}_M)$, $\boldsymbol{\Sigma}_{\boldsymbol{e}_1}=\dots =\boldsymbol{\Sigma}_{\boldsymbol{e}_N}=0.02\boldsymbol{I}_M$, $\hat{p}_1=\dots =\hat{p}_N=0.9$, and $\eta = 0.2$. The numerical results of connect probability, SER and transmit power against the SNR requirement are shown in Fig. \ref{f2}.

Fig. 2(a) shows the connect probability results of various schemes. It can be noted that NRob and Ref do not satisfy the connect probability requirement (as marked with dashed line), while SphB performs the over-satisfaction, and Iter-SphB performs the best approximation. The influence of the satisfaction of the connect probability requirement can be reflected in Fig. 2(b, c), the SER and transmit power results. With higher connect probability, NBob, SphB and Iter-SphB will perform better in SER, worse in transmit power, whereas Ref perform unstable (even worse than NRob with the high SNR requirement) due to incorrect derivation as mentioned before.

\section{Conclusion}

In this letter, we propose a sphere bounding scheme for probabilistic robust CI power minimizing precoding, and then propose the relaxation iteration for the scheme. Besides, we prove the lower bound of the tightened CI max-min SNR precoding can be solved through tightened CI power minimizing precoding. We finally analyze the complexity of our proposed schemes. Numerical results show that our proposed schemes perform well in the satisfaction of the connect probability requirement, resulting in lower SER and higher transmit power.


%

\appendices
\section{Proof of Lemma \ref{l1}}

Optimization problem \eqref{e17} can be equivalently rewritten as
\begin{subequations}\label{e20}
\begin{gather}  {\rm min}_{\tilde{\boldsymbol{x}}, t}\, -t \\
\text{s.t.}\quad  \|\tilde{\boldsymbol{x}}\|^2 \le \hat{P}, \label{e20_1} \\
\boldsymbol{E}_i^-\tilde{\boldsymbol{x}} - t\boldsymbol{f}_i \succeq_{M+1} \boldsymbol{0}, \boldsymbol{E}_i^+\tilde{\boldsymbol{x}} - t\boldsymbol{f}_i \succeq_{M+1} \boldsymbol{0}, i\in N, \label{e20_2} \end{gather}
\end{subequations}

\noindent where
$\boldsymbol{E}_i^{\mp} = [r_i\tilde{\boldsymbol{\Sigma}}_{\boldsymbol{e}_i}^{1/2}(\boldsymbol{A} \mp \boldsymbol{B}/{{\rm tan}\theta})\tilde{\boldsymbol{x}};\tilde{\boldsymbol{h}}_i^{\rm T}(\boldsymbol{A} \mp \boldsymbol{B}/{{\rm tan}\theta})]$,
$r_i = \sqrt{2}{\rm erf}^{-1}(\hat{p}_i)$,
$\boldsymbol{f}_i = [\boldsymbol{0}; \sigma_{z_i}]$, and
$\boldsymbol{u} = [u_1,\dots ,u_K]^{\rm T} \in \mathcal{R}^{K\times 1} \succeq_K \boldsymbol{0}$
represents that $\boldsymbol{u}$ belongs to SOC of dimension $K$ ($u_K \ge \sqrt{u_1^2 + \dots u_{K-1}^2}$). The Lagrangian of optimization problem \eqref{e20} can be expressed as
\begin{align}\label{e21}
\mathcal{L}(\tilde{\boldsymbol{x}}, t, \lambda_0, \boldsymbol{\lambda}_i^{\mp}) ={}& -t - \lambda_0(\hat{P} - \|\tilde{\boldsymbol{x}}\|^2) \notag \\
& - \sum\nolimits_{i=1}^N {(\boldsymbol{\lambda}_i^{\mp})^{\rm T}(\boldsymbol{E}_i^{\mp}\tilde{\boldsymbol{x}} - t\boldsymbol{f}_i)},
\end{align}

\noindent where the dual variables $\lambda_0 \ge 0$ and $\boldsymbol{\lambda}_i^{\mp} \succeq_{M+1} \boldsymbol{0}$. The KKT conditions can be obtained as
\begin{subequations}\label{e22}
\begin{align}
& \frac{\partial \mathcal{L}}{\partial \tilde{\boldsymbol{x}}} = 2\lambda_0 \tilde{\boldsymbol{x}} - \sum\nolimits_{i=1}^N {(\boldsymbol{E}_i^{\mp})^{\rm T} \boldsymbol{\lambda}_i^{\mp}}  = \boldsymbol{0}, \label{e22_1} \\
& \frac{\partial \mathcal{L}}{\partial t} = -1 + \sum\nolimits_{i=1}^N {(\boldsymbol{\lambda}_i^{\mp})^{\rm T} \boldsymbol{f}_i} = 0, \label{e22_2} \\
& \lambda_0(\|\tilde{\boldsymbol{x}}\|^2 - \hat{P}) = 0, \label{e22_3} \\
& (\boldsymbol{\lambda}_i^{\mp})^{\rm T}(\boldsymbol{E}_i^{\mp}\tilde{\boldsymbol{x}} - t\boldsymbol{f}_i) = 0, i \in N. \label{e22_4}
\end{align}
\end{subequations}

\noindent We have $\lambda_0 \ne 0$,
$\tilde{\boldsymbol{x}} = (\sum\nolimits_{i=1}^N {(\boldsymbol{E}_i^{\mp})^{\rm T} \boldsymbol{\lambda}_i^{\mp}})/(2\lambda_0)$, $\|\tilde{\boldsymbol{x}}\|^2 = \hat{P}$,
$\sum\nolimits_{i=1}^N {(\boldsymbol{\lambda}_i^{\mp})^{\rm T} \boldsymbol{f}_i} = 1$,
and $t = 2\lambda_0 \hat{P}$. The dual problem (strong duality is easily verified \cite{abental0001}) can be written as
\begin{subequations}\label{e23}
\begin{gather}  {\rm max}_{\lambda_0, \boldsymbol{\lambda}_i^{\mp}}\, -2 \lambda_0 \|\tilde{\boldsymbol{x}}\|^2\\
\text{s.t.}\quad  \tilde{\boldsymbol{x}} = (\sum\nolimits_{i=1}^N {(\boldsymbol{E}_i^{\mp})^{\rm T} \boldsymbol{\lambda}_i^{\mp}})/ (2\lambda_0), \|\tilde{\boldsymbol{x}}\|^2 = \hat{P}, \label{e23_1}\\
\lambda_0 > 0, \boldsymbol{\lambda}_i^{\mp} \succeq_{M+1} \boldsymbol{0}, i \in N, \label{e23_2}
\end{gather}
\end{subequations}

\noindent which can be relaxed as
\begin{subequations}\label{e24}
\begin{gather}  {\rm min}_{\tilde{\boldsymbol{x}}, \lambda_0}\, 2 \lambda_0 \|\tilde{\boldsymbol{x}}\|^2\\
\text{s.t.}\quad  \boldsymbol{E}_i^{\mp}\tilde{\boldsymbol{x}} - 2\lambda_0 \hat{P} \boldsymbol{f}_i \succeq_{M+1} \boldsymbol{0}, i \in N, \label{e24_1} \\
\|\tilde{\boldsymbol{x}}\|^2 = \hat{P}, \lambda_0 > 0. \label{e24_2}
\end{gather}
\end{subequations}

\noindent It takes the same form as optimization problem \eqref{e16}.


%
%

\ifCLASSOPTIONcaptionsoff
  \newpage
\fi

\end{document}